\documentclass[prr,superscriptaddress,twocolumn,longbibliography]{revtex4-1}

\usepackage{amsmath}
\usepackage{amssymb}
\usepackage{amsxtra,color}
\usepackage{latexsym}
\usepackage{graphicx}
\usepackage{tikz}
\usepackage{pgfplots}
\usepgfplotslibrary{groupplots}
\pgfplotsset{compat=1.3}
\usepackage{subcaption}
\usepackage{lipsum}
\usepackage{color}
\usepackage{braket}
\usepackage{lineno}
\usepackage{ulem}
\usepackage{wasysym}

\usepackage[utf8]{inputenc}
\usepackage{MnSymbol}	

\usepackage{hyperref}

\definecolor{MyDarkGreen}{rgb}{0,0.6,0}
\definecolor{MyDarkBlue}{rgb}{0,0,0.8}
\definecolor{MyDarkRed}{rgb}{0.6,0,0.3}

\begin{document}

\title{The influence of final state interactions in attosecond photoelectron interferometry}

\author{S.~\surname{Luo}}\thanks{These authors contributed equally}
\affiliation{Department of Physics, Lund University, Box 118, 22100 Lund, Sweden}

\author{R.~\surname{Weissenbilder}}\thanks{These authors contributed equally}
\affiliation{Department of Physics, Lund University, Box 118, 22100 Lund, Sweden}

\author{H.~\surname{Laurell}}
\affiliation{Department of Physics, Lund University, Box 118, 22100 Lund, Sweden}

\author{R.~Y.~\surname{Bello}}
\affiliation{Departamento de Química Física Aplicada, Módulo 14, Universidad Autónoma de Madrid, E-28049 Madrid, Spain}

\author{C.~\surname{Marante}} 
\affiliation{Department of Physics, University of Central Florida, Orlando, Florida 32816, USA}

\author{M.~\surname{Ammitzb{\"o}ll}}
\affiliation{Department of Physics, Lund University, Box 118, 22100 Lund, Sweden}

\author{L.~\surname{Neori\v{c}i\'c}}
\affiliation{Department of Physics, Lund University, Box 118, 22100 Lund, Sweden}

\author{A.~\surname{Ljungdahl}}
\affiliation{Department of Physics, Stockholm University, AlbaNova University Center, SE-106 91 Stockholm, Sweden}

\author{R.~J.~\surname{Squibb}}
\affiliation{Department of Physics, University of Gothenburg, Origov\"agen 6B, 41296 Gothenburg, Sweden}

\author{R.~\surname{Feifel}}
\affiliation{Department of Physics, University of Gothenburg, Origov\"agen 6B, 41296 Gothenburg, Sweden}

\author{M.~\surname{Gisselbrecht}}
\affiliation{Department of Physics, Lund University, Box 118, 22100 Lund, Sweden}

\author{C.~L.~\surname{Arnold}}
\affiliation{Department of Physics, Lund University, Box 118, 22100 Lund, Sweden}

\author{F.~\surname{Mart\'in}}
\affiliation{Departamento de Química, Módulo 13, Universidad Autónoma de Madrid, E-28049 Madrid, Spain}
\affiliation{Instituto Madrileño de Estudios Avanzados en Nanociencia (IMDEA-Nanociencia), Cantoblanco,E-28049 Madrid, Spain}

\author{E.~\surname{Lindroth}}
\affiliation{Department of Physics, Stockholm University, AlbaNova University Center, SE-106 91 Stockholm, Sweden}

\author{L.~\surname{Argenti}} 
\affiliation{Department of Physics, University of Central Florida, Orlando, Florida 32816, USA}

\author{D.~\surname{Busto}}
\affiliation{Department of Physics, Lund University, Box 118, 22100 Lund, Sweden}
\affiliation{Institute of Physics, Albert Ludwig University, Stefan-Meier-Strasse 19, 79104  Freiburg, Germany}

\author{A.~\surname{L'Huillier}}
\affiliation{Department of Physics, Lund University, Box 118, 22100 Lund, Sweden}
\email{anne.lhuillier@fysik.lth.se}

\pacs{}

\begin{abstract}

Fano resonances are ubiquitous phenomena appearing in many fields of physics, e.g. atomic or molecular photoionization, or electron transport in quantum dots. Recently, attosecond interferometric techniques have been used to measure the amplitude and phase of photoelectron wavepackets close to Fano resonances in argon and helium, allowing for the retrieval of the temporal dynamics of the photoionization process. In this work, we study the photoionization of argon atoms close to the $3s^13p^64p$ autoionizing state using an interferometric technique with high spectral resolution. The phase shows a monotonic 2$\pi$ increase across the resonance or a sigmo\"idal less than $\pi$ variation depending on experimental conditions, e.g. the probe laser bandwidth. Using three different, state-of-the-art calculations, we show that the measured phase is influenced by the interaction between final states reached by two-photon transitions. 
   
\end{abstract}

\maketitle

\section{Introduction}

In 1935, Beutler observed spectral lines with asymmetric absorption profiles~\cite{Beutler1935}. In a seminal article~\cite{FanoPR1961}, Fano explained this asymmetry by the superposition of a discrete state and a continuum coinciding in energy, interacting with each other and leading to an interference effect. The Beutler-Fano formula, which describes this interaction with only a few key parameters, has become a workhorse in many fields of physics, including nuclear, atomic, molecular, and condensed matter physics~\cite{KronerNature2008,MiroshnichenkoRMP2010,LimonovNP2017,PengNP2022}.

As derived by Fano~\cite{FanoPR1961}, quantum interference between direct atomic photoionization and ionization due to excitation of a discrete state that decays by autoionization can be described with the Fano resonance factor
\begin{equation}
		R(\epsilon)=\frac{q+\epsilon}{\epsilon+i},
  \label{eq1}
\end{equation}
which multiplies the transition dipole moment corresponding to direct photoionization. In this expression, $\epsilon=2(E-E_r)/\Gamma$ is the reduced energy, equal to zero at the position of the resonance and normalized by the resonance width $\Gamma$, and  $q$ is the asymmetry parameter, which depends on the relative strength between the transition from the ground state to the autoionizing state and direct photoionization. When $q$ is real, Eq.~(\ref{eq1}) leads to an asymmetric line shape due to interference effects between the different ionization paths and to an abrupt $\pi$ phase jump when $q+\epsilon=0$.

In the initial description of asymmetric lineshapes in (single) photon absorption, $q$ is a real quantity~\cite{FanoRMP1968,MaddenPRL1963,SchmidtRPP1992}. However, further studies showed that $q$ becomes complex-valued in above-threshold-ionization (e.g. two-photon transitions)~\cite{cormier1993Complex_q,sanchez1995Complex_q}, when the autoionizing state interacts with multiple channels~\cite{Mendoza2008_complexq}, or when multiple Fano resonances overlap~\cite{Magunov_complexq}. A complex-valued $q$-parameter is also needed to fully describe quantum systems with decoherence, for example electron transport through quantum dots~\cite{KobayashiPRL2002}, microwave transmission in metal cavities~\cite{AndreasPRL2010,MaPRL2022}, or atoms in the presence of a strong laser field~\cite{MarlenePRL2005,ZielinskiPRL2015}, and systems with time-reversal symmetry breaking~\cite{ClerkTRSbreaking,AndreasPRL2010}. 

Fano resonances have been extensively studied in atomic systems, especially noble gases, using synchrotron radiation~\cite{MaddenPRL1963,SchmidtRPP1992,SorensenPRA1994,DomkePRA1996}. In such experiments, the absorption profiles are measured with high spectral resolution, leading to a precise determination of $q$. Recently, the advent of laser-based high-order harmonic radiation in the extreme ultraviolet regime~\cite{McPherson,Ferray}, when combined with a probe laser field, has allowed new types of measurements, elevating our understanding and control of the interference phenomenon~\cite{WangPRL2010,GalanPRL2014,KoturNC2016,grusonScience2016,BustoJPB2018,BarreauPRL2019,TurconiJPB2020,BustoEPJD2022}. In some of these experiments, the spectral phase and amplitude of the autoionizing wave packet was measured using an interferometric technique (RABBIT, reconstruction of attosecond beating by interference of two-photon transitions)~\cite{KoturNC2016,grusonScience2016,BustoJPB2018,BarreauPRL2019,TurconiJPB2020,BustoEPJD2022}. As a consequence, the ultrafast temporal dynamics associated with a Fano resonance~\cite{grusonScience2016,BustoEPJD2022} could be reconstructed.

An important question is how much the probe laser field perturbs the interaction and modifies the resonance profile. Laser-induced changes in the spectral profiles have been studied at relatively high laser intensities in experiments~\cite{WangPRL2010,OttScience2013}, and in simulations~\cite{MarlenePRL2005}. The results could be described by the introduction of a complex-valued $q$ parameter~\cite{ZielinskiPRL2015,AgarwalPRA1984}. Even in a perturbative regime, as in RABBIT measurements, the probe field might induce additional interaction channels, affecting the interferometric measurements.

In the present work, we study the photoionization of argon atoms using a high spectral resolution attosecond interferometric setup~\cite{LuoSetup2023}. The laser wavelength is tuned so that the 17th harmonic, generated in an argon gas cell, is resonant with the quasi-bound $3s^{-1}4p$ state. Measurements are performed using the energy-resolved (``rainbow'') RABBIT technique~\cite{grusonScience2016}, based on the use of a weak infrared (IR) probe field. The main experimental novelty compared to previous measurements~\cite{KoturNC2016,TurconiJPB2020} is the use of a narrow-bandwidth (10\,nm wide) probe field. This allows us to separate the two spin--orbit components, which are found to show similar phase and amplitude behavior. Depending on the measurement, performed in sideband 16 (SB16, one IR photon below the resonance) or SB18 (one IR photon above the resonance) the phase is found to vary by almost $2\pi$, or much less, about $1.4$\,rad, across the resonance. The strong $2\pi$ phase variation was not observed in previous measurements~\cite{KoturNC2016,TurconiJPB2020}. We also investigate the dependence of this phase variation with probe bandwidth and pump wavelength. Using calculations, we show that final state interactions involving (non-resonant) dipole transitions in the $3s3p^6n\ell$ discrete spectrum followed by autoionization to the final  $3s^23p^5\epsilon\ell'$ continuum states strongly affect the measurements.

The article is structured as follows. In Section~II, we describe the experimental method and results. Theoretical methods and calculations are presented in Section~III. In Section~IV, 
we discuss the role of the final state interactions on the measured phase variation. We conclude in Section~V.

\begin{figure*}[t]
    \centering
   \includegraphics[width=0.8\linewidth]{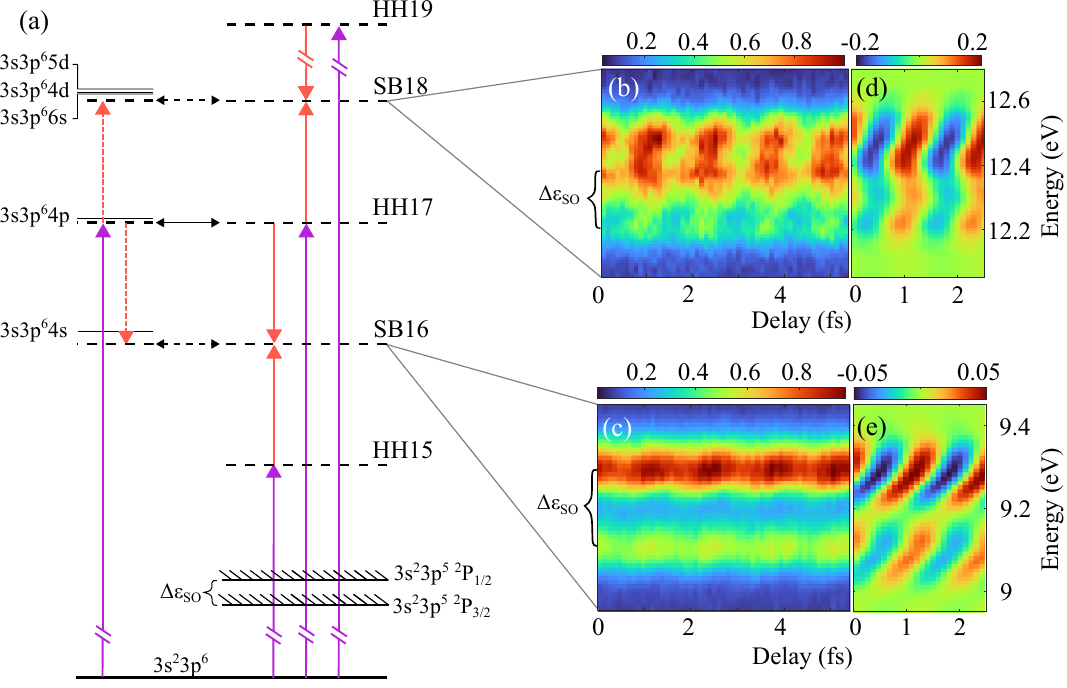}
    \caption{(a) Illustration of the levels, channels, and processes involved in the experiment. The purple (red) arrows represent the harmonic (IR) photons. The black arrows indicate autoionization. Dashed lines mark continuum or virtual states reached by absorption of a harmonic or a harmonic $\pm$ an IR photon. Solid lines are quasi-bound states which may decay by autoionization. (b) Measured photoelectron spectra in SB18 and (c) in SB16 in argon as a function of delay between XUV and IR.  
    (d) Photoelectron spectra in SB18 and (e) in SB16 after subtracting the mean values of the oscillations.}
    \label{RABBITscanning}
\end{figure*}
\section{Experimental method and results}

\subsection{Experimental setup}

A 4\,mJ, 40\,fs infrared (IR) titanium sapphire laser pulse is split into a pump and a probe pulse in a Mach-Zehnder interferometer~\cite{LuoSetup2023}. The pump pulse is focused into a gas cell filled with argon to generate high-order harmonics. A 200\,nm thick aluminum filter blocks the remaining pump IR pulse and transmits the generated XUV pulses. The probe pulse is sent to a variable-delay piezo stage to enable the variation of delay between pump and probe pulses. Its bandwidth can be reduced from  35\,nm to 10\,nm by adding an interference filter. The XUV pump and IR probe pulses are recombined on a holey mirror, and are finally focused by a toroidal mirror into an argon gas jet located in the interaction region of a magnetic bottle electron spectrometer (MBES). The experiment consists in measuring the photoelectron spectra, integrated over all angles, as a function of the pump-probe delay. 

The experiments presented in this work rely on high spectral resolution achieved through the use of a narrow-band probe pulse. In addition, photoelectrons are detected with a MBES with a 2\,m-long time-of-flight tube, allowing an energy resolution better than 50\,meV for SB16 and 90\,meV for SB18. As a consequence of the good spectral resolution, the spin--orbit structure is well resolved, for the first time in this type of measurement~\cite{KoturNC2016,TurconiJPB2020}.  High temporal stability, estimated to be 15\,as or 0.07\,rad, is achieved by active stabilization of the Mach-Zehnder interferometer~\cite{LuoSetup2023}. 

\subsection{Energy levels and transitions}

Figure~\ref{RABBITscanning}(a) illustrates the different energy levels and transitions involved in the present work. Argon atoms are exposed to XUV radiation (purple arrows) in the form of odd high-order harmonics (labelled HH$q$) of an IR laser field, and a narrowband IR probe laser (red arrows) with a variable delay relative to the XUV. Absorption of one harmonic, HH$_q$, plus absorption of an IR photon, and absorption of the next harmonic, HH$_{q+2}$, and emission of an IR photon, therefore, lead to the same final state (sidebands, denoted SB16 or SB18 in the figure).

Ionization of argon following a one or two-photon transition leads to photoelectrons whose energy depend on the final ionic state, $3p^{5}$ $^{2}P_{1/2}$ and $3p^{5}$ $^{2}P_{3/2}$, due to the spin--orbit interaction. The two spin--orbit components are separated by 177\,meV and the (statistical) ratio between the two is close to 1/2. 

The left part of Fig.~\ref{RABBITscanning}(a) shows excited states $3s3p^6n\ell$, (also noted $3s^{-1}n\ell$), which can be reached by absorption in the $3s^2$ subshell of one harmonic photon ($3s3p^64p$) or one harmonic plus/minus one IR photon ($3s3p^64s$, $4d$, $6s$ or $5d$). Their energies are given in Table~\ref{tab:experiment}. These excited states are embedded in the continuum and decay by autoionization. The $3s3p^64p$ state may decay either by autoionization, or by absorption/emission of an IR photon followed by autoionization. The IR intensity used in the experiments is kept low enough so that the probability to absorb or emit more than one IR photon is negligible.

\begin{table}[t]
\centering
\caption{Energy levels of resonant states in argon. ${^a}$Taken from Ref.~\cite{lin2003optically}. ${^b}$Taken from Ref.~\cite{wu1995electron}.}
\setlength{\tabcolsep}{2mm}{
\begin{tabular}{cccccc} \toprule
States& $3s^{-1}4s$ & $3s^{-1}4p$ & $3s^{-1}4d$ & $3s^{-1}6s$ & $3s^{-1}5d$\\ \hline
$E_{r}$(eV) &25.21${^{a}}$& 26.60${^b}$& 28.27${^a}$& 28.31${^a}$&28.63${^a}$\\\hline
$\Gamma$(meV) &202${^a}$& 76${^b}$& 5${^a}$&  & 5${^a}$\\
\hline
\end{tabular}}
\label{tab:experiment}
\end{table}

\subsection{Experimental method}

The laser wavelength is tuned so that the 17th harmonic is close to the $3s^{-1}4p$ autoionizing state. Consequently, both SB16 and SB18 can be reached by a path influenced by the Fano resonance, involving absorption of the 17th harmonic, and another path, by absorption of the 15th harmonic for SB16 or the 19th harmonic for SB18.  

Figure~\ref{RABBITscanning}(b,c) presents measured photoelectron spectra as a function of delay between the XUV and IR pulses in SB18 and SB16.  The sideband peaks present clear oscillations, with a period equal to half a laser cycle. The spin--orbit splitting is well resolved, which is an improvement compared to previous measurements~\cite{KoturNC2016,TurconiJPB2020}. For each energy, the sideband signal can be described as 
\begin{equation}
    S(\tau) = |A^{+}|^2 +|A^{-}|^2 + 2|A^{+}||A^{-}|\cos(2\omega\tau-\Delta\varphi),
    \label{eq2}
\end{equation} 
where $A^{+}$ and $A^{-}$ are the energy-dependent two-photon transition amplitudes involving absorption ($+$) or emission ($-$) of an IR photon and $\Delta\varphi= \varphi^{-}-\varphi^{+}$ is the phase difference. 
The sideband oscillation is analyzed by fitting Eq.~(\ref{eq2}) to the experimental data for each energy. The delay-dependent photoelectron spectra, with the mean values of the oscillations removed, are shown in Fig.~\ref{RABBITscanning}(d,e) for SB18 and SB16, respectively. 

\subsection{Experimental results}
\begin{figure*}[t]
    \centering
   \includegraphics[width=0.9\linewidth]{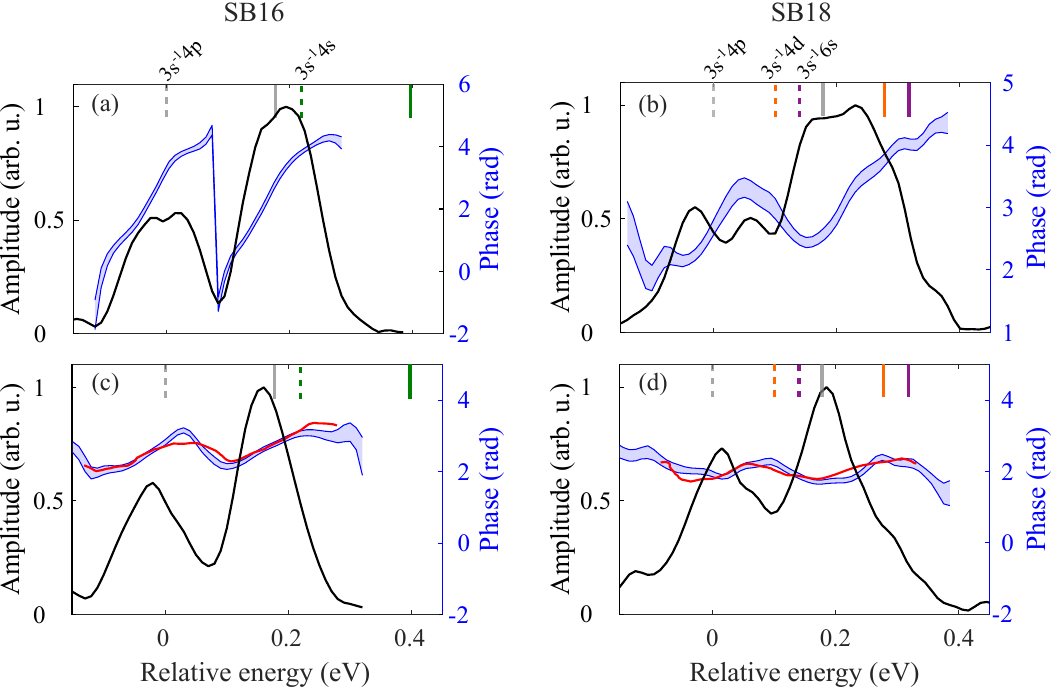}
    \caption{Amplitude (black lines) and phase 
(blue lines) of (a) SB16 and (b) SB18 measured with a 10\,nm probe bandwidth, and of (c) SB16 and (d) SB18 measured with a 35\,nm probe bandwidth. The red lines are adapted from Ref.~\cite{TurconiJPB2020}, with a shift for comparison. The vertical lines indicate the position of the $3s^{-1}4s$ (green), $3s^{-1}4d$ (orange), $3s^{-1}6s$ (violet) resonances, relative to the position of the $3s^{-1}4p$ resonance plus/minus one IR photon (grey) for SB18 and SB16, in the spectrum corresponding to the $^2P_{1/2}$ (dashed) $^2P_{3/2}$ (solid) ionic state,}
    \label{AmpPhase_Bandwidth}
\end{figure*}

In Fig.~\ref{AmpPhase_Bandwidth}(a,b), we present the phase and amplitude variation as a function of energy in SB16 and SB18. To facilitate comparisons between the experimental results and the different calculations presented in Sec.~III, we present kinetic energies relative to $E_r\pm\hbar\omega-I_\mathrm{p}(^2P_{1/2})$, where $E_r$ is the energy of the $3s^{-1}4p$ resonance and where $\pm$ refers to the IR photon being emitted (-) or absorbed (+). The positions of resonances accessible through two-photon transitions are indicated by vertical lines at the top of each panel in the figure. The phase (blue line) is obtained by a fit of a cosine function [see Eq.~(\ref{eq2})], to the experimental data. Both amplitude and phase variations show similar results for the two spin--orbit components, with a shift in energy corresponding to the spin--orbit splitting (177\,meV) and an amplitude ratio close to 1/2. No significant phase difference between the two spin--orbit components can be observed. 

The amplitude variations in Fig.~\ref{AmpPhase_Bandwidth}(a,b) exhibit minima at the resonance position characteristic of a window resonance. For SB16, the phase variation across the resonance is almost $2\pi$, which is at variance with previous measurements of the same process~\cite{KoturNC2016,TurconiJPB2020}, while for SB18, it is around 1.4\,rad. Note that the sharp phase jump between the two spin--orbit components in Fig.~\ref{AmpPhase_Bandwidth}(a) is a non-physical $2\pi$ phase jump. In contrast, the phase of SB20 (not shown) is approximately flat, with a slight increase of 0.1\,rad per 0.1\,eV.

We indicate in Fig.~\ref{AmpPhase_Bandwidth} the position of the quasi-bound states. The $3s^{-1}4p$ shifted by plus or minus one IR photon is indicated in grey, where the dashed (solid) line indicates the position of the shifted resonance corresponding to the $^2P_{1/2}$ ($^2P_{3/2}$) final ionic state. The other resonances that can be reached from the $3s^{-1}4p$ by one-photon absorption or emission are shown by the colored lines. While the $3s^{-1}4s$ is not directly resonant with the final continua reached in SB16, in SB18, the $3s^{-1}4d$ and $3s^{-1}6s$ states are quasi-resonant.
 
\subsection{Influence of the probe bandwidth and pump wavelength}
To understand better the difference between the measurements and previous results reported in the literature\cite{KoturNC2016,TurconiJPB2020}, we performed complementary experiments by investigating the influence of the probe bandwidth on the phase variation of SB16 and SB18 across the resonance. Figure~\ref{AmpPhase_Bandwidth}(c,d) presents results obtained in similar conditions as in Fig.~\ref{AmpPhase_Bandwidth}(a,b), with the difference being the bandwidth of the probe field, now equal to 35\,nm. As expected, the resonant features are smoothed out compared to the measurements with a 10~nm bandwidth probe pulse. The $2\pi$ phase variation is reduced to less than 1\,rad, consistent with previous measurements of the same resonance using a 50\,nm probe bandwidth~\cite{TurconiJPB2020}, shown in red. In SB18, the phase variation is also reduced when using shorter probe pulses and reproduces well the results in Ref.~\cite{TurconiJPB2020}.

\begin{figure*}
    \centering
   \includegraphics[width=0.8\linewidth]{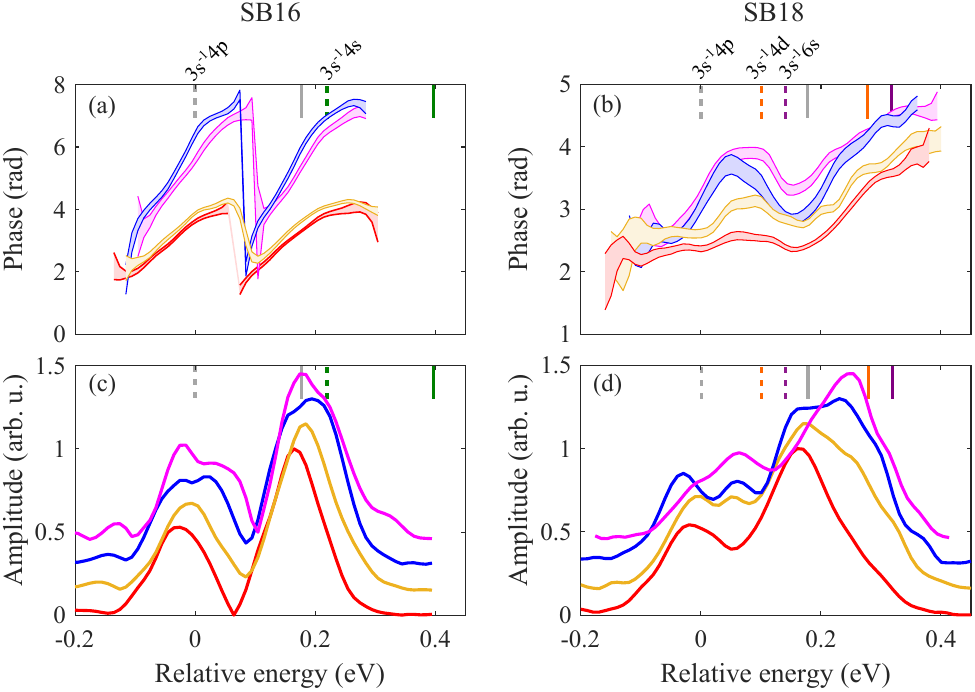}
    \caption{Phase and amplitude of SB16 (a,c), and SB18 (b,d) for different pump IR wavelengths (788\,nm in magenta, 789\,nm in blue, 790\,nm in yellow, and 791\,nm in red). The probe wavelength is 790\,nm. For visibility, a vertical shift is added in the amplitudes between the results obtained at different wavelengths. The vertical lines indicate the positions of the $3s^{-1}4s$ (green), $3s^{-1}4d$ (orange), $3s^{-1}6s$ (purple) resonances, relative to the position of the $3s^{-1}4p$ resonance plus/minus one IR photon (grey) for SB18 and SB16, in the spectrum corresponding to the $^2P_{1/2}$ (dashed) and $^2P_{3/2}$ (solid) ionic state.}
    \label{wavelengthScan}
\end{figure*}

Finally, we investigated the influence of the pump wavelength on the amplitude and phase variation of SB16 and SB18. In these measurements, the probe central wavelength and bandwidth, imposed by the transmission of the interference filter, are fixed to 790\,nm and 10\,nm, respectively. The results are shown in Fig.~\ref{wavelengthScan}(a,c) for SB16 and Fig.~\ref{wavelengthScan}(b,d) for SB18. The phase variation is strongly affected by the central wavelength of the pump. Between the results at 788\,nm and 789\,nm, and those at 790\,nm and 791\,nm, the phase variation of SB16 changes from almost $2\pi$ to less than $\pi$. The phase variation of SB18 is also more pronounced at 788\,nm and 789\,nm. The amplitudes of SB16 and SB18 change, with a clear indication of the $3s^{-1}4p$ window resonance at 788\,nm and 789\,nm, and narrower peaks at 790\,nm and 791\,nm.  

The photon energy of HH17 decreases as the pump laser is tuned from 788\,nm to 791\,nm. Consequently, the energy of SB16 approaches the $3s^{-1}4s$ resonance, leading to a change in the phase variation, thus emphasizing the influence of final state interactions. As shown in Table~\ref{tab:experiment}, the states involved in SB18 via two-photon transitions, in particular $3s^{-1}4d$ and $3s^{-1}6s$, exhibit significantly narrower linewidths than the $3s^{-1}4s$ state. Interestingly, the change in pump wavelength also affects the measured phase variation, in particular for the $^2P_{1/2}$ final ionic state as seen in Fig.~\ref{wavelengthScan}(b).

\section{Theoretical interpretation}

To interpret the experimental findings, we have employed three different and independent computational approaches. The first approach utilizes many-body perturbation theory to describe two-photon ionization processes.
While this method omits finite pulse effects and yields resonance positions that do not align with the experimental values, making a direct comparison with the experiment difficult, its ability to selectively switch on or off interactions is very useful for discussing the importance of the different physical effects. In the second approach, we use XCHEM, an {\it ab initio} molecular ionization program suite based on the close-coupling method~\cite{Marante2017}, to simulate the photoionization process with finite pulses. Thanks to its large-scale configuration interaction approach, XCHEM can account for most of the short-range correlation and the inter-channel coupling within the electrostatic approximation. However, XCHEM cannot account for spin--orbit splitting, and cannot be used for pulses as long as those used in the experiment. To address these limitations, our third approach employs NewStock~\cite{Carette2013,Cariker2021}, an atomic-ionization close-coupling code. While XCHEM can include a larger number of configurations, NewStock can semi-empirically account for spin--orbit interactions and simulate photoionization events with longer pulses. Table~\ref{tab:theory} shows the energy and width of the $3s^{-1}4p$ and $3s^{-1}4s$ states of the three computational methods. The positions of the resonances calculated using RPAE are obtained by using the experimental value of the $3s$ ionization energy instead of the Hartree-Fock value. The resonance widths cannot be accurately estimated using RPAE since higher order correlation, involving the $3p^{-2}3dn\ell$ configurations, are not included. The positions of the $^1S$ and $^1D$ resonances were further adjusted slightly to match the experimental values shown in Table~\ref{tab:experiment}.

\begin{table}[t]
\centering
\caption{Theoretical energy and width of the autoionizing states relevant to SB16, computed with three different numerical approaches.}
\setlength{\tabcolsep}{2mm}{
\begin{tabular}{ccccc} \toprule
& \multicolumn{2}{c}{$3s^{-1}4s$} & \multicolumn{2}{c}{$3s^{-1}4p$} \\ \hline
 & $E_{r}$(eV) & $\Gamma$(meV) & $E_{r}$(eV) & $\Gamma$(meV)  \\
\hline
RPAE & 25.44 & 12 & 26.74 & 11 \\
\hline
XCHEM & 25.46 & 114.24 & 26.73 & 73\\
\hline
NewStock & 24.91 & 138 & 26.07 & 64  \\
\hline
\label{tab:theory}
\end{tabular}}            
\end{table}

\subsection{Analysis using many-body perturbation theory}

\begin{figure*}[t]
    \centering
\includegraphics[width=0.85\linewidth]{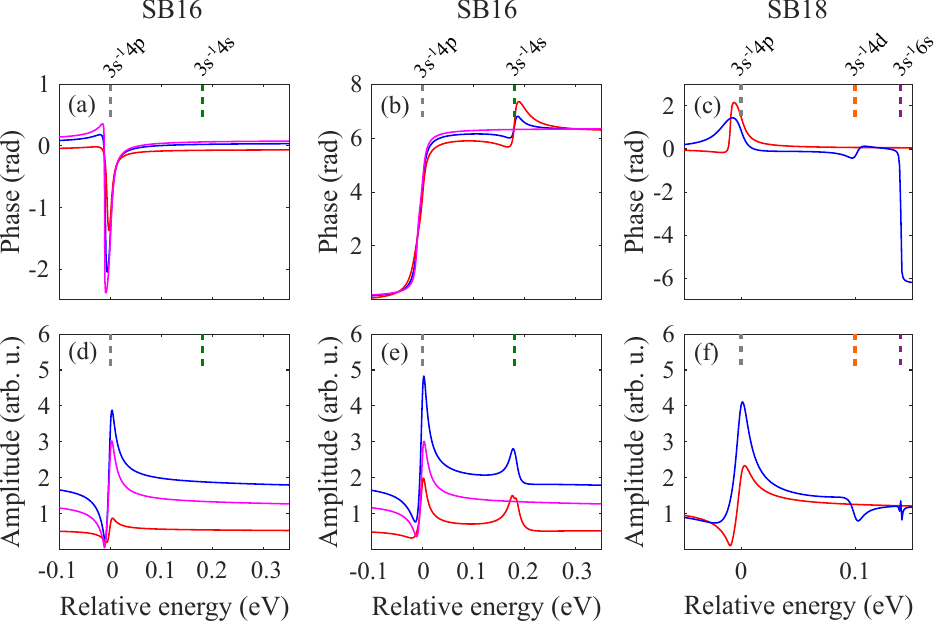}
    \caption{Phase and amplitude of two-photon matrix elements calculated using the RPAE approach. Phase and amplitude of SB16 (a,d) without and (b,e) with final state correlations included (blue lines). The red (magenta) lines corresponds to the phase and amplitude of the $\epsilon p$ ($\epsilon f$) final state. (c) Phase and (f) amplitude of SB18. The red and blue lines indicate the results without and with final state interactions, respectively. The vertical dashed lines indicate the position of the $3s^{-1}4s$ (green), $3s^{-1}4d$ (orange), $3s^{-1}6s$ (purple) resonances, relative to the position of the $3s^{-1}4p$ resonance plus/minus one IR photon (grey) for SB18 and SB16, respectively.}
    \label{RPAE}
\end{figure*}

In the many-body perturbation theory approach, interaction with the probe field is accounted for to lowest-order, i.e. only one-photon absorption or stimulated emission is allowed for. The two-color two-photon random phase approximation with exchange (RPAE) approach~\cite{vinbladh2019} is used to calculate correlated two-photon matrix elements that include channel coupling, after both one- and two-photon interaction, as well as ground state correlation. We investigated the influence of final state correlations via the interaction of the $3sns$ and $3snd$ states with the final continua, thus including the paths indicated by the dashed red and black arrows in Fig.~\ref{RABBITscanning}(a). 

The phase and amplitude in SB16 without these final state correlations are indicated by the blue lines in Fig.~\ref{RPAE}(a,d). This corresponds to only including interactions described by solid (red and black) arrows in Fig.~\ref{RABBITscanning}(a), i.e. autoionization in the vicinity of the $3s^{-1}4p$ state, and IR photon absorption or stimulated emission in the $3s^23p^5$ spectrum. The red and magenta lines show the phase and amplitude of the $3s^23p^5\epsilon p$ and $3s^23p^5\epsilon f$ final states, respectively. In the $\epsilon f$ final state, the amplitude shows perfect destructive interference and hence a sharper, and larger, phase variation across the resonance than in the $\epsilon p$ final state. The reduced contrast of the resonance in the $\epsilon p$ final state is due to the two contributing paths to the final state (autoionization to the $\epsilon s$ and $\epsilon d$ continua, followed by stimulated emission of an IR photon). 

The phase and amplitude of SB16, including final state correlations, are shown in Fig.~\ref{RPAE}(b,e). The results are obtained by including also IR photon emission in the $3s3p^6$ spectrum, followed by autoionization to the final $3s^23p^5$ spectrum [dashed red and black arrows in Fig.~\ref{RABBITscanning}(a)]. These transitions involve both the bound part of the $3s^{-1}4p$ resonance, and the non-resonant contribution, which, however, is expected to contribute less. These results show the importance of the final state interactions for the measurement of the phase variation across the $3s^{-1}4p$ resonance. In the case of SB16, comparing Fig.~\ref{RPAE}(a) and Fig.~\ref{RPAE}(b), the phase changes from first decreasing then increasing as the energy increases, with a total phase change of less than $\pi$, to a monotonic $2\pi$ phase increase. Additionally, when plotting separately the contributions of the two final states $3s^23p^5\epsilon p$ and $3s^23p^5\epsilon f$ [red and magenta lines in Fig.~\ref{RPAE}(b,e)], the influence of the $4s$ resonance is only observed in the $3s^23p^5\epsilon p$ final continuum, due to angular momentum coupling rules. Both cases, however, exhibit a $2\pi$ phase variation across the $3s^{-1}4p$ resonance. This shows that final state interactions, which are certainly enhanced by the proximity
of the $3s^{-1}4s$ state for the $3p\rightarrow\epsilon p$ channel, also influence the $3p\rightarrow\epsilon f$ channel (possibly through the non-resonant contributions of the $3s^{-1}nd$ quasi-bound states).

In SB18 [Fig.~\ref{RPAE}(c)], the phase without including final state interactions (red line) is very similar to that of SB16 [Fig.~\ref{RPAE}(a)], except for the expected sign change~\cite{KoturNC2016}. In contrast, when including final state interactions in SB18 (blue line), the result is quite different from SB16. This emphasizes the importance of final state interactions, which are different in the two spectral regions (for SB16 and SB18). In SB16, the $3p \rightarrow \epsilon p$ channel might be influenced by  the broad $3s^{-1}4s$ resonance, while for SB18, several narrow resonances, e.g. $3s^{-1}6s$, or $3s^{-1}4d$ might be important.

For the amplitude, the difference between the calculations performed with and without final state interactions is much less pronounced than for the phase, both in SB16 and SB18. In the case of SB16, the amplitude variation is dominated by the Fano resonance in the intermediate, one-photon transition. The influence of narrow resonances in the final state can be observed in the amplitude variation of SB18. The main effect of adding final state interactions is to reduce the interference contrast (incomplete destructive interference), especially for SB18.

\subsection{Studying the influence of the probe photon energy and pulse duration using XCHEM}

\begin{figure*}[t]
    \centering
\includegraphics[width=0.8\linewidth]{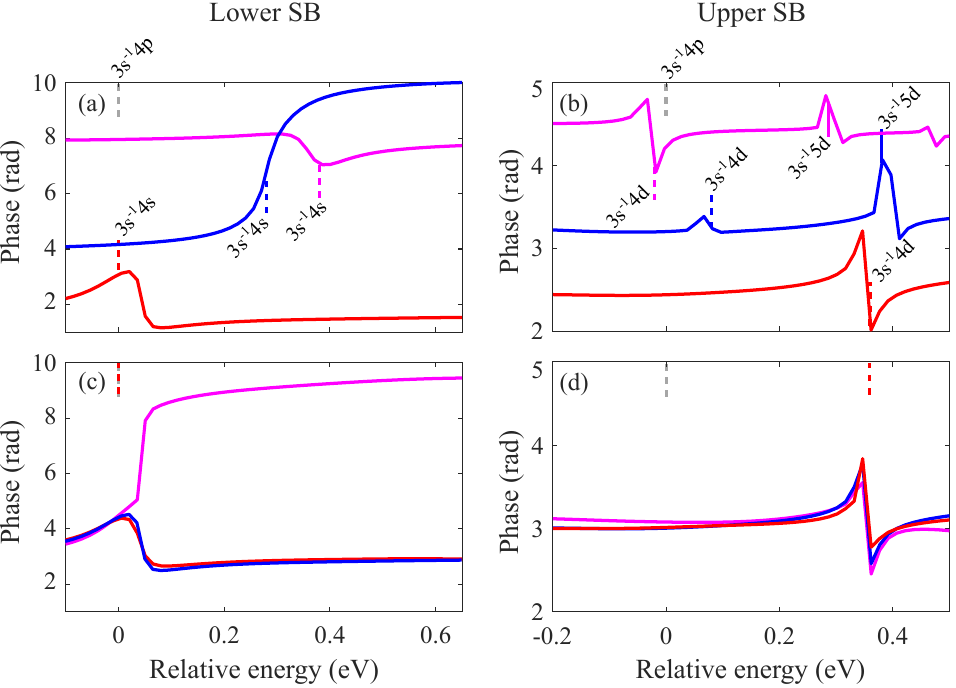}
    \caption{Phase of (a) lower sideband and (b) upper sideband as a function of the probe photon energy [1.27\,eV (red), 1.55\,eV (blue), 1.65\,eV (magenta)] obtained using the XCHEM approach. The probe pulse duration is 12\,fs. The dashed vertical lines in (a) indicate the position of the $3s^{-1}4s$ relative to the position of the $3s^{-1}4p$ resonance minus one IR photon (grey) for the three different probe energies, and in (b) the dashed (solid) vertical lines correspond to the $3s^{-1}4d$ ($3s^{-1}5d$) resonance relative to the position of the $3s^{-1}4p$ resonance plus one IR photon (grey). Phase of (c) the lower sideband and (d) the upper sideband as a function of pulse duration [8\,fs (red), 12\,fs (blue) and 22\,fs (magenta)]. The probe photon energy is 1.27\,eV. The dashed red vertical line indicates the position of the $3s^{-1}4s$ resonance in (c), and $3s^{-1}4d$ resonance in (d), relative to the position of the $3s^{-1}4p$ resonance plus/minus one IR photon (grey) for the upper and lower sideband, respectively.
    }
    \label{XCHEM}
\end{figure*}

We now present the results obtained using the XCHEM approach. The ground state of argon ($^1\!S^e$) was obtained from a complete active space configuration interaction (CAS-CI) calculation, including the first four $s$, three $p$ and one $d$ orbitals, with the $1s$, $2s$ and $2p$ core orbitals always doubly occupied. These orbitals were optimized using the state-average Restricted Active Space SCF(SA-RASSCF) capability of MOLCAS~\cite{molcas} where two $^1\!S^e$ states were included in the state average. We used the aug-cc-pVTZ one-electron basis set~\cite{Dunning_1989,Dunning_1992}. The cation states were obtained by constructing the (\textit{N}-1)-electron configuration state functions (CSFs) using the same orbitals as in the MOLCAS calculation of the ground state. The three components of the $^2\!P$ state of the cation (energetically open) and the $^2\!S^e$ state of the cation (closed channel) were included in the close-coupling calculation, with $\ell_{max}=6$ for the partial wave expansion of the scattering states. The TDSE was solved by expanding the time-dependent wave function in the basis of fully-correlated electronic states obtained in the previous steps~\cite{Bello_2021}. We note that a similar basis of electronic states was used in previous work to obtain one- and two-photon single ionization cross sections of argon in very good agreement with previous experimental and theoretical results~\cite{Bello2022}.

In Fig.~\ref{XCHEM}(a,b) we present the energy-resolved phase of the sidebands extracted from simulated RABBIT traces using probe photon energies of 1.27\,eV (red), 1.55\,eV (blue) and 1.65\,eV (magenta), for pulses with a duration of 12\,fs. The simulation with a probe energy of 1.27\,eV is chosen to exactly match the energy difference between the $3s^{-1}4p$ and $3s^{-1}4s$ states. The ionizing XUV pulse is prepared in such a way that one of the harmonics is in resonance with the $3s^{-1}4p$ state. We note that the sideband orders used in the XCHEM calculations are different from those used in the experiment due to the difference in photon energy. In the following we refer to the upper and lower sidebands relative to the position of the $3s^{-1}4p$ resonance.


For all three probe photon energies, the phase of the lower sideband features a sudden variation in the vicinity of the $3s^{-1}4s$ resonance, while for photon energies equal to 1.55\,eV and 1.65\,eV it is relatively flat close to the $3s^{-1}4p$ resonance. The absence of a large phase variation in the vicinity of the $3s^{-1}4p$ resonance is due to the use of very short IR probe pulses (shorter than in the experiments) which significantly smooths out the phase variation. In contrast, the phase variation in the vicinity of $3s^{-1}4s$ resonance is not affected by finite pulse effects since it is directly resonant with the final continua.  As the probe photon energy increases from 1.27\,eV to 1.55\,eV, the phase variation changes from a sigmo\"idal shape featuring less than $\pi$ variation, to a monotonic 2$\pi$ phase variation. Increasing the probe photon energy to 1.65\,eV, the phase variation reverts to a sigmo\"idal less than $\pi$ variation, of different sign. The 2$\pi$ phase variation is present for photon energies varying from 1.55\,eV to 1.57\,eV, which corresponds to the probe photon energy used in the experiment. The wavelength-dependent calculations indicate that the participation of both the $3s^{-1}4p$ and $3s^{-1}4s$ resonances in the two-photon transition leads to the 2$\pi$ phase variation observed in the lower sideband, since it disappears both for a probe energy of 1.27\,eV, when $3s^{-1}4s$ dominates, and for a probe energy of 1.65\,eV, when the two-photon transition is dominated by the $3s^{-1}4p$ resonance. In the upper sideband, the phase variation shows the signature of multiple final state resonances, while, as for the lower sideband, finite pulse effects smooth out the contribution of the intermediate $3s^{-1}4p$ resonance. Due to the very narrow width of the final state resonances the phase variation is not well resolved.

Figure~\ref{XCHEM}(c,d) shows the phase of the lower and upper sidebands for a probe photon energy of 1.27\,eV and pulse durations of 8\,fs, 12\,fs and 22\,fs. The phase variation is strongly affected by the probe bandwidth. As the pulse duration increases and the bandwidth decreases, the phase variation switches from a sigmo\"idal variation to a 2$\pi$ phase increase in the lower sideband. In contrast, in the upper sideband only small phase variations are observed, which, due to the relatively broad probe bandwidth and several close lying quasi-bound states, are dominated by the two-photon resonances ($3s^{-1}6s$, $3s^{-1}4d$, and $3s^{-1}5d$). Additionally, there is significant overlap between the $3s^{-1}6s$ and $3s^{-1}4d$ resonances, further reducing the observed phase variation.

\subsection{Simulations including spin--orbit interaction}

\begin{figure*}
\centering
\includegraphics[width=0.8\linewidth]{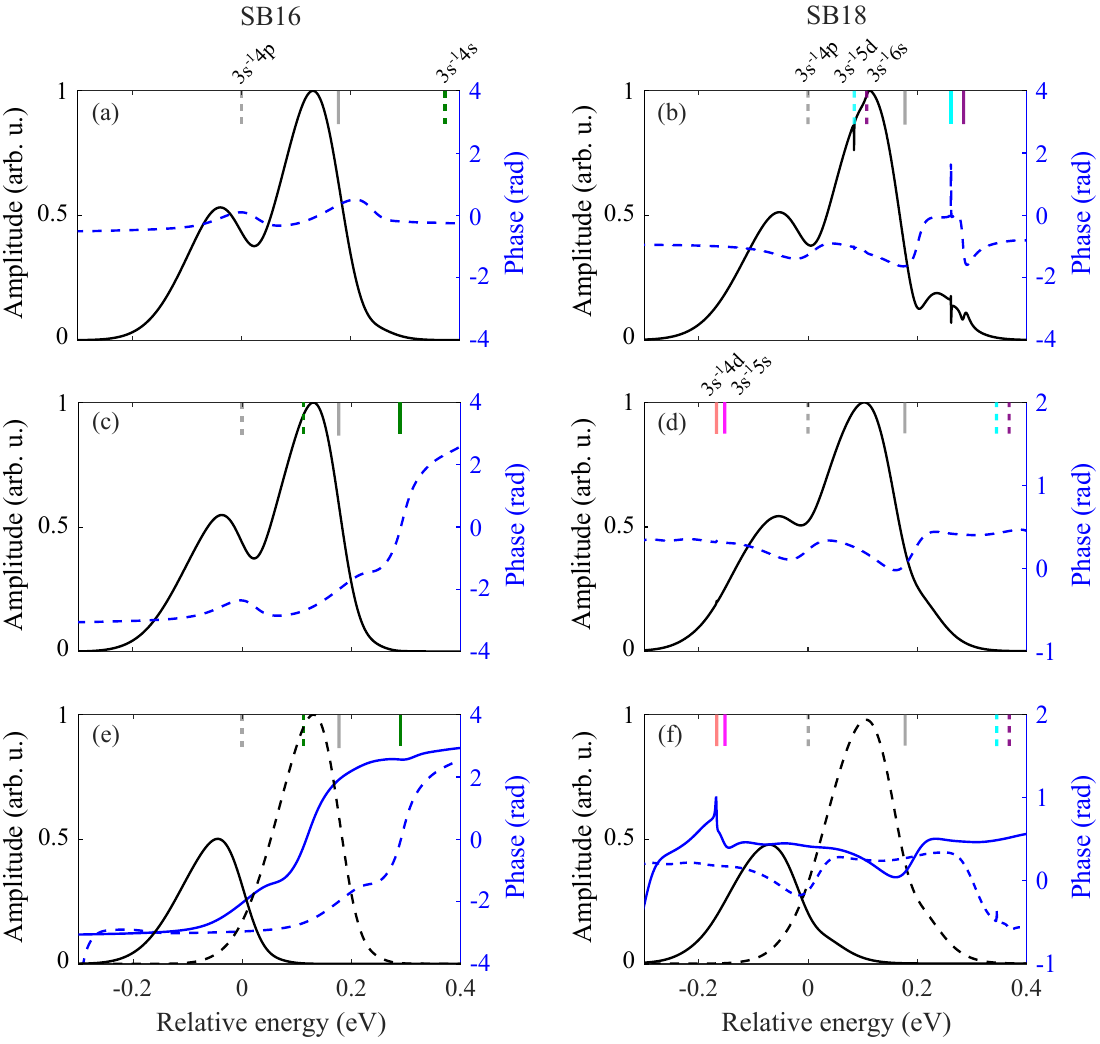}
    \caption{Amplitude (black lines) and phase (blue lines) of the RABBIT beating around (a,c,e) SB16 and (b,d,f) SB18, extracted from the theoretical RABBIT spectrum computed with NewStock, including the spin--orbit coupling. In (a,b) the IR probe pulse has a duration of 46\,fs and a central wavelength of 809\,nm. In (c,d,e,f) the IR probe pulse has a duration of 55\,fs and a central wavelength of 975\,nm. In both cases the IR has an intensity of  0.1\,$\mathrm{TW/cm^{2}}$. In (e,f) the amplitude and phase of the individual contributions from the $^2\!P_{1/2}$ and $^2\!P_{3/2}$ ionic channels are shown, indicated by the solid and dashed lines, respectively. The vertical lines indicate the position of the $3s^{-1}4s$ (green), $3s^{-1}5d$ (cyan), $3s^{-1}5s$ (magenta), $3s^{-1}4d$ (orange) and $3s^{-1}6s$ (purple) resonances, relative to the position of the $3s^{-1}4p$ resonance plus/minus one IR photon (grey) for SB18 and SB16, in the spectrum corresponding to the $^2P_{1/2}$ (dashed) and $^2P_{3/2}$ (solid) ionic state.
    }
    \label{fig:NewStock}
\end{figure*}

Finally, we simulated the photoionization of argon using the NewStock atomic photoionization code~\cite{Carette2013}, extended to account for the spin--orbit interaction. To a first approximation, the relativistic effects in the photoionization of argon from its valence shell can be reduced to the spin--orbit splitting of the $^2\!P_{1/2,3/2}$ parent ion. To reproduce these effects, we recouple the non-relativistic states obtained in \textit{ab initio} calculations to give rise to ionization states with well-defined total angular momentum $J$, and we add to the non-relativistic Hamiltonian a $J$-dependent energy shift for the matrix elements between states originating from the same ion with total angular momentum $J$. The XUV attosecond-pulse train comprises three consecutive harmonics, HH15, HH17, and HH19, with a FWHM of approximately 11\,fs, and the IR probe pulse is varied between 46\,fs and 55\,fs. These values are chosen to mimic the conditions of some of the experiments. The central energy of HH17 is selected to be resonant with the theoretical value of the $3s^{-1}4p$ autoionizing state energy, 26.07\,eV. 

Figure~\ref{fig:NewStock}(a,b) shows the energy-resolved phase and amplitude of SB16 and SB18 of the simulated energy-resolved RABBIT photoelectron spectrum computed with a pulse duration of 46\,fs. The parameters of the oscillations are directly obtained from the complex amplitudes of the two contributing two-photon transitions to the sideband. The peaks in the theoretical photoelectron spectrum are slightly shifted with respect to the experiment due the relative error in the absolute energy of the computed autoionizing states, and to the laser wavelength which is larger than in the experiment in order to match the detuning of SB16 from the $3s^{-1}4s$ autoionizing state. With this set of parameters, the simulations reproduce well the experimental conditions in Fig.~\ref{AmpPhase_Bandwidth}(c,d), where the experimental phase for SB16 does not exhibit a $2\pi$ jump. Indeed, for SB16, the theoretical results quantitatively reproduce both the proportion of the two spin--orbit split peak amplitudes, as well as the associated phase shifts. SB18 comprises several autoionizing states with even symmetry, which complicates the spectrum. Nevertheless, even for this sideband, we find a good qualitative agreement in both the amplitude and phase of the main resonant features. 

To compensate for the discrepancy between the experimental and {\it ab initio} energies of the $3s^{-1}4s$ and $3s^{-1}4p$ states, we repeated the simulations using a 975\,nm IR pulse. In this way, the center of SB16 is closer to the energy of the $3s^{-1}4s$ autoionizing state, as it is in the experiment. Furthermore, to approximate the longer IR pulse duration used in the experiments in which the $2\pi$ jump is observed, we have also used a longer pulse, with a duration of 55\,fs. Figure~\ref{fig:NewStock}(c,d) shows the amplitude and phase of SB16 and SB18 extracted from this simulation, alongside the amplitude and phase of the contributions to the sideband signal from each of the two ionic channels ($^2\!P_{3/2}$ and $^2\!P_{1/2}$). When the signals of the two ionic channels are considered individually in SB16, as presented in blue dashed lines in Fig.~\ref{fig:NewStock}(c), the phase of each of the two channels exhibits a $2\pi$ variation at the position of the $3s^{-1}4s$ resonance. When the two ionic signals are added incoherently, however, only the second $2\pi$ variation survives, whereas the phase variation in the $^2\!P_{1/2}$ channel is reduced to a small bump. We note that this is similar to recent spin--orbit resolved RABBIT simulations of the $3s^{-1}4p$ resonance, using the time-dependent R-matrix method~\cite{Roantree2023}, where each ion-resolved RABBIT signal exhibits a close to $\pi$ phase jump, whereas the joint spectrum shows a strongly reduced phase variation in the $^2\!P_{1/2}$ spin--orbit component. Similarly, in SB18 [Fig.~\ref{fig:NewStock}(d)], the total phase follows the isolated $^2\!P_{3/2}$ component closely, while showing a reduced phase variation compared to the isolated $^2\!P_{1/2}$ component.

\section{Discussion}

\begin{figure*}[ht!]
\centering
\includegraphics[width=0.7\linewidth]{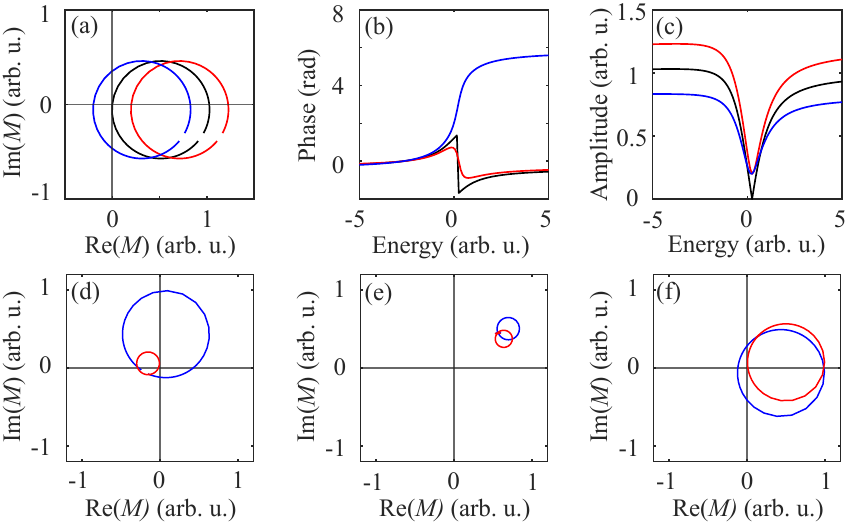}
\caption{(a) Complex representation of $M(\epsilon)$ without background (black), with a positive background (red) and with a negative background (blue). (b) $\arg[M(\epsilon)]$ and (c) $|M(\epsilon)|^2$ as a function of $\epsilon$ for the three cases. (d-f) Channel resolved complex trajectory of the two-photon matrix element in the vicinity of SB16 calculated with RPAE with the final state (d) $\epsilon p\,^1\!S$, (e) $\epsilon{p}\,^1\!D$ and (f) $\epsilon{f}\,^1\!D$. The red (blue) lines indicate the results without (with) final state interactions. The amplitudes of the complex trajectories in (d-f) have been normalized to the respective maximum radius in each panel.}
\label{fig:circles}
\end{figure*}


A common feature of all the experimental and theoretical results presented in the previous sections is the extreme sensitivity with wavelength and pulse duration of the phase variation across the $3s^{-1}4p$ resonance, which can be as large as $2\pi$ or close to zero depending on the chosen sideband.
A $2\pi$ phase variation across a Fano resonance has been shown theoretically to occur under certain conditions when the asymmetry parameter acquires a non-zero imaginary component~\cite{GalanPRA2016}. There are several processes that can lead to a change from the purely real $q$-parameter described in Fano's original paper, to a complex one. Two such processes are inherent to the case of two-photon ionization of argon atoms studied here. First, in general, when a two-photon transition involving an autoionizing state (in the intermediate or final state) occurs, it results in an effective $q$-parameter with a non-zero imaginary part~\cite{cormier1993Complex_q,sanchez1995Complex_q,GalanPRA2016,vzitnik2022Complex_q}. The strength of the radiative coupling between the bound state and the continuum [dashed red and black arrows in Fig.~\ref{RABBITscanning}(a)] influences the magnitude of the imaginary part of $q$. When the two-photon transition is in the vicinity of a resonance, the coupling strength, and thus the effect on the measured phase, can be strongly enhanced. 

Secondly, as discussed above, for the $3s^{-1}4p$ autoionizing resonance there are several angular channels through which the state can decay, leading to an effective complex $q$, even in single-photon ionization~\cite{Mendoza2008_complexq}. In this case, the interaction of the bound state with the final $s$ and $d$ continua can be transformed into an interacting, and a non-interacting continuum channel~\cite{KoturNC2016}, where the non-interacting channel only adds a smoothly varying background term to the two-photon transition matrix element, as
\begin{equation}
		M(\epsilon)=A_{0}\frac{q+\epsilon}{\epsilon+i}+A_{bg}=(A_{0}+A_{bg})\frac{\tilde{q}+\epsilon}{\epsilon+i},
  \label{eq3}
\end{equation}
with $\tilde{q} = (A_{0}q+iA_{bg})/(A_{0}+A_{bg})$, where $A_0$ and $A_{bg}$ are the amplitudes of the interacting and non-interacting channels, respectively.

To illustrate how an effective complex $q$, resulting from either of the above mentioned processes, can influence the phase and amplitude measurements, we show in Fig.~\ref{fig:circles}(a) the complex trajectory of a transition matrix element, with $q=-0.25$, describing a window resonance such as $3s^{-1}4p$ in argon~\cite{SorensenPRA1994}, in the case of zero background (purely real $q$, black line), and when including a small (real) positive or negative background (red and blue line). Figure~\ref{fig:circles}(b,c) shows the corresponding phase and amplitude variation in the three cases. In the case of no background term, the complex trajectory intersects the origin, leading to a sharp $\pi$ phase variation, and an amplitude showing perfect destructive interference due to the window resonance. For the positive background, the circle describing the complex trajectory shifts to the right, so that it no longer intersects the origin. The phase variation shows a similar trend as for the real $q$, but with a reduced, smoother phase variation. Finally, when adding a negative background, the complex trajectory encloses the origin, which leads to a smooth $2\pi$ phase variation across the resonance.   

As shown in a previous work~\cite{GalanPRA2016}, taking into account finite pulse (bandwidth) effects in the description of resonant two-photon ionization, the circle describing the variation of the complex amplitude with energy shrinks as the bandwidth increases, due to mixing of different spectral components. In our {experimental conditions, the trajectory encloses the origin for SB16 when using a narrow 10\,nm probe bandwidth, while as the bandwidth is increased to 35\,nm, the circle becomes smaller and no longer encloses the origin. Thus the phase variation changes from a monotonic increase of almost 2$\pi$ [see Fig.~\ref{AmpPhase_Bandwidth}(a)] to a sigmo\"idal variation of less than $\pi$ [see Fig.~\ref{AmpPhase_Bandwidth}(c)]. A similar effect is observed in the XCHEM and NewStock calculations reported in the previous section.


To complement this simple model, we show the complex trajectory of the two-photon matrix elements in SB16, calculated using RPAE, for the $\epsilon p\,^{1}\!S$, $\epsilon p\,^{1}\!D$ and $\epsilon f\,^{1}\!D$ final states in Fig.~\ref{fig:circles}(d-f). The red (blue) circles indicate calculations without (with) final state correlations. As can be seen, including final state correlation shifts the circle to enclose the origin for the $\epsilon\,p ^{1}\!S$ and $\epsilon f \,^{1}\!D$ final states, leading to a $2\pi$ phase variation in these cases. In addition, especially for the $\epsilon p\,^1\!S$  final state, the radius is increased. For the $\epsilon p\,^1\!D$ final state, however, the red circle is far away from the origin, indicating a stronger background contribution, due to the contribution of the intermediate $s$ and $d$ continua, leading to a small amplitude and phase variation across the $3s^{-1}4p$ resonance. The effect of final state correlations on the phase and amplitude is relatively small in this case. Similar representations of the two-photon matrix elements calculated with XCHEM and NewStock are discussed in Appendix A.



\section{Conclusion}
In summary, we have studied resonant two-photon ionization via the $3s^{-1}4p$ autoionizing state in argon, using rainbow RABBIT with a high-resolution photoelectron spectrometer. The autoionizing state is excited by absorption of the 17th harmonic and phase measurements are performed by interferometry both in sideband 16 and in sideband 18. The high spectral resolution allows us to resolve the two spin--orbit components, for which the measured spectral amplitude and phase are observed to be very similar. The phase variation observed in SB16 depends on the probe bandwidth, being almost 2$\pi$ using a narrow 10\,nm bandwidth, and less than $\pi$ using a 35\,nm bandwidth. The phase variation is much less pronounced in SB18, independently of the probe bandwidth. The phase variation is also affected by the pump wavelength.

We attribute these results to the influence of final state interactions, through the absorption or emission of an additional photon from the quasi-bound state to an energy that in some cases is close to a resonant state. The photon absorption or emission is then followed by autoionization. This additional path interferes with the usual RABBIT paths.  The effect of these processes on the measured Fano profile and phase can be understood by representing the trajectory described by the complex amplitude. As a result of channel mixing and final state interactions the asymmetry parameter $q$ is modified and becomes complex valued causing a shift of the trajectory in the complex plane. Small shifts do not significantly modify the amplitude but can dramatically affect the phase variation going from a smaller than $\pi$ sigmo\"{i}dal phase variation to a monotonously increasing 2$\pi$ phase excursion.

Theoretical results based on many-body perturbation theory~\cite{vinbladh2019}, the XCHEM approach~\cite{Marante2017}, and the NewStock atomic photoionization code~\cite{Carette2013}, extended to account for the spin--orbit interaction, illustrate different aspects of the physics studied in the present work. Apart from the asymmetry between SB16 and SB18, a strong dependence on probe wavelength and pulse duration is observed. In a recent theoretical work~\cite{Roantree2023}, RABBIT simulations in the vicinity of the $3s^{-1}4p$ resonance in argon were investigated using semi-relativistic time-dependent R-matrix calculations. Their results confirm the absence of significant differences in the amplitude and phase associated with the two spin--orbit components. Interestingly, the authors also observe a strong dependence of the phase of SB16 as a function of the IR driving wavelength.

Two main conclusions can be drawn from the experimental and theoretical studies presented here. (i) Phase measurements performed with high spectral resolution are extremely sensitive to the details of the interaction and can be used to test advanced theoretical models. (ii)  Final state interactions involving absorption or emission of one IR photon in the $3s3p^6n\ell$ spectrum followed by autoionization play an important role. This implies that the use of RABBIT measurements to determine the ``phase of a Fano resonance'' should be examined carefully for each case. In the study presented here, because of final state correlation effects, the measured phase in SB16 or SB18 differs from the phase of the $3s3p^64p$ resonance, excited by absorption of XUV-only radiation.

\section*{Acknowledgements}

The authors acknowledge support from the Swedish Research Council (2013-8185, 2016-04907, 2018-03731, 2020-0520, 2020-03315, 2020-06384, 2023-04603), the European Research Council (advanced grant QPAP, 884900) and the Knut and Alice Wallenberg Foundation. This work was developed in the framework of the COST action CA18222 Attosecond Chemistry (AttoChem) supported by European Cooperation in Science and Technology. AL and MA are partly supported by the Wallenberg Center for Quantum Technology (WACQT) funded by the Knut and Alice Wallenberg Foundation. LA and CM acknowledge the NSF grant No. PHY-1912507. RYB and FM acknowledge the Ministerio de Ciencia e Innovaci\'{o}n MICINN (Spain) through projects PID2022-138288NB-C31 and PID2022-138288NB-C32, the Severo Ochoa Programme for Centres of Excellence in R\,$\&$\,D (CEX2020-001039-S) and the Mar\'{i}a de Maeztu Programme for Units of Excellence in R\,$\&$\,D (CEX2018-000805-M). XCHEM  calculations were performed at the Marenostrum  Barcelona Supercomputer Center of the Red Espa\~nola de Supercomputaci\'on and the Centro de Computaci\'{o}n Cient\'{i}fica of Universidad Aut\'{o}noma de Madrid.

\appendix
\section{Complex trajectories of two-photon matrix elements calculated using XCHEM and NewStock}

\begin{figure}[hb!]
\centering
\includegraphics[width=0.9\linewidth]{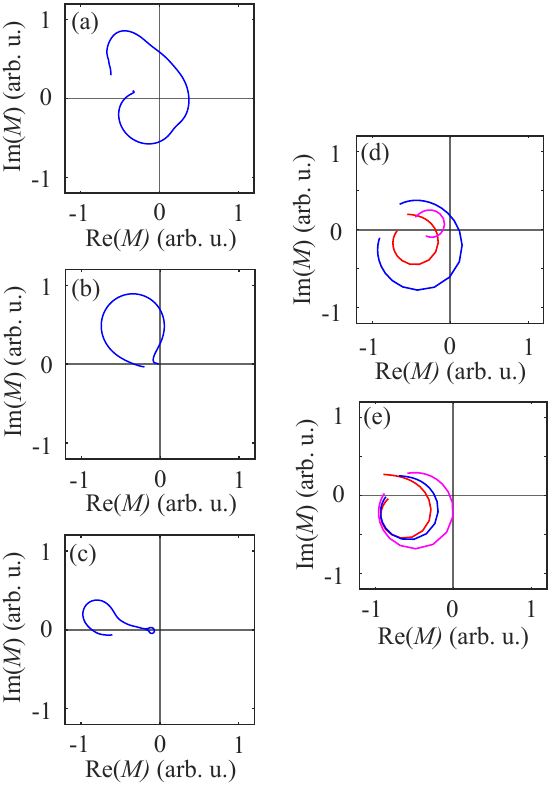}
\caption{(a-c) Channel resolved complex trajectory of the two-photon matrix element in the vicinity of SB16 calculated with NewStock with the final state (a) $^2\!P_{1/2}\,\epsilon p_{1/2}\,^1\!S$, (b) $^2\!P_{1/2}\,\epsilon p_{3/2}\,^1\!D$  and (c) $^2\!P_{1/2}\,\epsilon f_{5/2}\,^1\!D$. (d-e) Complex trajectory of the coherent superposition of the $\epsilon{p}\,^1\!S$ and $\epsilon{p}\,^1\!D$ two-photon matrix elements in the lower sideband, calculated with XCHEM. In (d) the probe pulse duration is $12\,$fs and the probe photon energy is $1.27\,$eV (red), $1.55\,$eV (blue) and $1.65\,$eV (magenta). In (e) the probe photon energy is $1.27\,$eV and the probe pulse duration is $8\,$fs (red), $12\,$fs (blue) and $22\,$fs (magenta). The amplitudes of the complex trajectories have been normalized to the respective maximum radius in each panel.}
\label{fig:circlesAppendix}
\end{figure}

It is interesting to compare the RPAE calculations to those obtained from NewStock and XCHEM. To obtain an estimate of the complex trajectories in the monochromatic limit, as a function of energy, from the finite pulse NewStock and XCHEM calculations, we normalize the amplitude of the resonant two-photon element (absorption of HH17 and stimulated emission of an IR photon) by the non-resonant amplitude (absorption of HH15 plus absorption of an IR photon). In Fig.~\ref{fig:circlesAppendix}(a-c) we show these estimated complex trajectories of the two-photon matrix elements, obtained with NewStock, as a function of energy, for the $\epsilon p\,^1\!S$, $\epsilon p\,^1\!D$ and $\epsilon f\,^1\!D$ final states. Due to the normalization procedure, the complex trajectories are distorted for energies far from the resonance, however, we still observe that the complex trajectory almost fully encloses the origin in the $\epsilon p\,^1\!S$ final state, implying a close to $2\pi$ phase variation. Contrary to the RPAE calculations, for the $\epsilon f\,^1\!D$ final state the complex trajectory is shifted to negative values and does not enclose the origin. As a result, there is no $2\pi$ phase variation across the resonance. This may indicate that a higher degree of correlation is needed to observe the $2\pi$ phase variation in this case.

In Fig.~\ref{fig:circlesAppendix}(d) and (e) we show the complex trajectory of the two-photon matrix element, corresponding to the coherent superposition of the $\epsilon{p}\,^1\!S$ and $\epsilon{p}\,^1\!D$ channels, obtained with XCHEM, for different probe photon energies and different probe pulse durations, respectively. Due to the nature of the XCHEM code the $\epsilon{p}\,^1\!S$ and $\epsilon{p}\,^1\!D$ channels cannot be separated, so a direct comparison to the RPAE and NewStock calculations is not possible, however, based on the RPAE calculations, the contribution of the $\epsilon{p}\,^1\!S$ channel is expected to be dominant. The complex trajectory in Fig.~\ref{fig:circlesAppendix}(d) is seen to encloses the origin only when a 1.55\,eV probe photon energy and 12\,fs probe pulse duration is used, leading to a $2\pi$ phase variation [see Fig.~\ref{XCHEM}(a)]. The effect of a changing probe pulse duration is shown in Fig.~\ref{fig:circlesAppendix}(e) for the case of a 1.27\,eV probe photon energy. When decreasing the pulse duration from 22\,fs to 12\,fs or 8\,fs the circular trajectories contract, leading to a smaller observed phase variation. When looking only at the coherent superposition of the $\epsilon{p}\,^1\!S$ and $\epsilon{p}\,^1\!D$ channels, the circle does not enclose the origin, even for a 22\,fs probe pulse duration, and as a result no $2\pi$ phase variation is expected. The $2\pi$ phase variation shown in Fig.~\ref{XCHEM}(c) is only observed when also the $\epsilon{f}\,^1\!D$ channel (not shown) is included. While the amplitude of the $\epsilon{f}\,^1\!D$ is small, and it does not display a $2\pi$ phase variation on its own, it is large enough to shift the circle to encompass the origin in the case of a 22\,fs probe pulse duration.

\bibliography{references.bib}

\end{document}